\documentclass[10pt,a4paper]{iopart}
\usepackage{iopams}  
\usepackage{graphicx,psfrag,bbm,latexsym,color,dcolumn,bm,dsfont,bbm,color,
mathrsfs,bbold,latexsym,amsfonts,amssymb}
\newcommand{\mediau}[1]{\int d\mu~ #1}
\newcommand{\media}[1]{\int d\mu_b~ #1}
\newcommand{\mediaT}[1]{\left\langle #1 \right\rangle}

\begin{document}


\title[Ising spin glass models versus Ising models II]
{Ising spin glass models versus Ising models: \\
an effective mapping at high temperature II.\\
Applications to graphs and networks}
%[A replica approach for the Ising spin glass model in zero field]
%{...the short-range spin glass model in zero field}

\author{Massimo Ostilli$^{1,2}$}
\address{$^1$\ Departamento de Fisica,
Universidade de Aveiro,
Campus Universitario de Santiago 3810-193, Aveiro, Portugal.}
%\address{$^1$\ Dipartimento di Fisica, Universit\`a di Roma ``La Sapienza'', 
%Piazzale A. Moro 2, Roma 00185, Italy}
\address{$^2$\ Center for Statistical Mechanics and Complexity, Istituto 
Nazionale per la Fisica della Materia, 
Unit\`a di Roma 1, Roma 00185, Italy.}

\ead{massimo.ostilli@roma1.infn.it}

\date{\today}

\begin{abstract}
By applying a recently proposed mapping,
we derive exactly the upper phase boundary of several Ising spin glass
models defined over static graphs and random graphs, 
generalizing some known results and providing new ones. 
\end{abstract}

%Spin Glass, Statistical Mechanics, Random Systems

\maketitle

\section{Introduction}
The science of graphs and networks is nowadays a well studied topic
receiving more and more interest due to its large
number of applications in the understanding of complex systems
in, technological, biological and social structures.
We refer the interested reader to the literature 
\cite{Barbasi,Dorogovtsev,Vespignani}.
As a basic starting point in the analysis of the cooperative behavior
typical of such complex systems, the study of the simplest 
dichotomy-interaction models, such as the Ising model, plays a crucial role
and in recent years successful efforts have been made in solving this model
\cite{Goltsev,Zecchina}, which, on networks, and especially on the so called
``scale free networks'', may manifest critical phenomena very different
from those over regular lattice systems \cite{Goltsev1,Goltsev2}.  
The Ising model with only ferromagnetic couplings, 
however, allows only a ``friendly'' interaction
between the elements of the network, $J_b=+J$, 
so that only a possible two-state
``friendly'' collective behavior is allowed.  
For taking into account that the interaction between the elements
of the network can also be unfriendly, $J_b=\pm J$, the Ising spin glass model
should instead be considered \cite{Kim}.

In the part I of this work \cite{I}, we have introduced 
a general mapping between a given Ising spin glass model
and a related Ising one;
the related Ising model, which, roughly speaking, is the model in which the
couplings are uniform. The mapping turns out to be of immediate application
when the solution of this related Ising model is known.

The mapping becomes exact when the
dimensionality of the system goes to infinite.
More precisely, we have seen that: for a system 
having a number of first neighbors per vertex proportional to $D$, 
the mapping applies exactly
as $D\to\infty$ (\textit{infinite dimensional in a strict sense})
in all the paramagnetic region;
whereas, more in general, the mapping becomes
an exact method to find the upper critical surface
%when, for any vertex, the probability that two randomly chosen
%infinitely long paths passing through the vertex, 
when, in the thermodynamic limit, 
the total number of infinitely long paths per vertex 
goes to infinite and, choosing randomly two of them,
the probability that they overlap each other 
for an infinite number of bonds goes to zero
(\textit{infinite dimensional in a large sense}).
%the number of infinitely long paths per vertex goes to infinite
%and almost all these paths are two at two independent
%(infinite dimensional in the large sense);
%two paths being independent if they share only a finite number of bonds. 

In \cite{I} we have applied the mapping to
several models in which the above condition is satisfied, including,
generalized Sherrington-Kirkpatrick models, which are infinite
dimensional in a strict sense, and the spin glass model
defined over the Bethe lattice, which is infinite dimensional
in the large sense. 

In this paper we will apply the mapping to general graphs and networks
which are infinite dimensional in the large sense. First, we will consider 
generalized static tree-like structures allowing also for a finite number 
of loops per vertex. Second, we will analyze the case of 
random graphs where an ensemble of graphs $G$, together with
a probability $P(g),~g\in G$, are given and 
all the graphs $g$ are infinite dimensional in the large sense, 
which, in particular, includes
the so called ``equilibrium random graphs'' \cite{Dorogovtsev}.
In both the cases, we find very general formulae for the exact 
upper critical surface: parmagnetic-spin glass and paramagnetic-disordered
ferromagnetic/antiferromagnetic transitions.
Whereas for the generalized tree-like structures these critical surfaces are
here for the first time given, we observe that, for the random graphs, the
critical surfaces have been just derived, but confined to  
a particular subclass of the equilibrium random graphs \cite{Kim}.
Furthermore, beyond the equilibrium random graphs,
we provide a formula which applies exactly
in the most general case for networks whose bonds 
are each other independent random variables.
This formula gives us the upper critical surface of the Ising spin glass model
in terms of the critical temperature of the Ising model over the network,
which, in general, has to be determined numerically, but, 
as a non random model, its measure turns out to be hugely easier 
than a direct measure for the spin glass model.

The paper is organized as follows. In Secs. 2 and 3, we recall briefly the
models we are considering and the main result of the Ref. \cite{I}.
In Sec. 4, we apply the mapping to static generalized tree-like structures.
In Sec. 5, a brief introduction for defining the random 
graphs is given. In the subsection 5.1, we recall the critical 
temperature of an Ising model defined over an ensemble of 
random graphs, which, by using the mapping, 
will be used in the following subsection 5.2 
to easily find the upper critical surface of a generic
Ising spin glass over an ensemble of random graphs.
Finally, some conclusions are reported in Sec. 6.

\section{Models}

\label{model}
In this paper we will apply the mapping to systems defined over a graph 
$g$ of $N$ vertices whose set of links $\Gamma_g$ will be
defined through the adjacency matrix of the graph, $g_{i,j}=0,1$:
\begin{eqnarray}
\label{Gamma1}
\Gamma_g\equiv \{b=\left(i_b,i_b\right): i_b,j_b \in g,
~ g_{i_b,j_b}=1,~ i_b<j_b\}.
\end{eqnarray} 
The fully connected graph will be indicated with $\Gamma_f$:
\begin{eqnarray}
\label{Gammaf}
\Gamma_f\equiv \{b=\left(i_b,i_b\right): i_b,j_b=1,\ldots,N, ~ i_b<j_b\}.
\end{eqnarray} 

The Hamiltonian of these spin glass models can be written as  
\begin{eqnarray}
\label{H}
%H_{\bm{J}}\left(\{\tilde{\sigma}_b\}\right)\equiv 
H\left(\{\sigma_i\};\{J_b\}\right)
\equiv -\sum_{b\in\Gamma_g} J_b \tilde{\sigma}_b
\end{eqnarray} 
where the $J_b$'s are 
quenched couplings, and $\tilde{\sigma}_b$ stays for the product
of two Ising variables, $\tilde{\sigma}_b=\sigma_{i_{b}}\sigma_{j_{b}}$, with
$i_{b}$ and $j_{b}$ such that $b=\left(i_b,j_b\right)$. 
%Note that such Ising variables are no more each other independent.

The free energy $F$ is defined by
\begin{eqnarray}
\label{logZ}
-\beta F\equiv \int d\mathcal{P}\left(\{J_b\}\right)
\log\left(Z\left(\{J_b\}\right)\right),
\end{eqnarray} 
where $Z\left(\{J_b\}\right)$ 
is the partition function of the quenched system
\begin{eqnarray}
\label{Z}
Z\left(\{J_b\}\right)= \sum_{\{\sigma_b\}}e^{-\beta 
H\left(\{\sigma_i\};\{J_b\}\}\right)}, 
\end{eqnarray} 
and $d\mathcal{P}\left(\{J_b\}\right)$ 
is a product measure given 
in terms of normalized measures $d\mu_b\geq 0$ 
(we are considering a general measure $d\mu_b$ 
allowing also for a possible dependence on the bonds) 
\begin{eqnarray}
\label{dP}
d\mathcal{P}\left(\{J_b\}\right)\equiv \prod_{b\in\Gamma_f} 
d\mu_b\left( J_b \right),
\quad \int d\mu_b\left( J_b \right) =1.
\end{eqnarray}
We take the Boltzmann constant $K_B=1$. 
A generic inverse critical temperature of the spin glass model, if any, 
will be indicated with $\beta_c$.

\section{Mapping}
In the part I of this work we have provided a mapping
between an Ising spin glass model and a related Ising one.
The mapping is exact on and above the critical temperature when,
in the thermodynamic limit, the dimensionality is infinite.
For the aims of the present work, we will need to recall
only the following definitions and rules of the mapping.

Given a spin glass model trough Eqs. (\ref{Gamma1}-\ref{dP}), 
we define, on the same set of links $\Gamma_g$, 
its \textit{related Ising model} 
trough the following Ising Hamiltonian
\begin{eqnarray}
\label{HI}
%H_{I}\left(\{\tilde{\sigma}_b\}\right)\equiv 
%-\sum_b J_b^{(I)} \tilde{\sigma}_b,
H_I\left(\{\sigma_i\};\{J_b\}\right)
\equiv -\sum_{b\in\Gamma_g} J_b^{(I)} \tilde{\sigma}_b
\end{eqnarray} 
where the Ising couplings $J_b^{(I)}$ have 
non random values such that $~\forall ~b,b'\in \Gamma_g$
%\numparts
\begin{eqnarray}
\label{JI}
J_{b'}^{(I)}&=&J_b^{(I)} \quad \mathrm{if} \quad 
d\mu_{b'}\equiv d\mu_{b}, \\
\label{JIb}
J_b^{(I)}&\neq & 0 \quad \mathrm{if} \int d\mu_b(J_b)J_b\neq 0 \quad
\mathrm{or} \quad \int d\mu_b(J_b)J_b^2>0. 
\end{eqnarray}
%\endnumparts
In the following, a suffix $I$ 
%over quantities such as $H_{I}$,
%$F_{I}$, $f_{I}$, etc\ldots, or $J_b^{(I)}$, $\beta_c^{(I)}$, etc\ldots,
will be referred to the related Ising system with Hamiltonian (\ref{HI}).
Finally, with $z_b$ ($z_{b}^{(I)}$) we will indicate the universal parameters
$z_b\equiv \tanh(\beta J_b)$ ($z_{b}^{(I)}\equiv \tanh(\beta J_{b}^{(I)})$),
and with $w_b$ ($w_{b}^{(I)}$) their critical value for $\beta_c$ 
($\beta_c^{(I)}$).

\subsection{Case of a uniform measure (same disorder for any bond)}
If $d\mu_b\equiv d\mu$ for any bond $b$ of $\Gamma_g$,
the related Ising model corresponds to a uniform Ising model
having a single coupling $J_b^{(I)}\equiv J^{(I)}$
and its critical behavior will be characterized by, at most, two points
$w_{F}^{(I)}=\tanh(\beta_{F}^{(I)}J^{(I)})>0$ and 
$w_{AF}^{(I)}=\tanh(\beta_{AF}^{(I)}J^{(I)})<0$, if any,
where $\beta_{F}^{(I)}$ and $\beta_{AF}^{(I)}$ are
the critical ferro- and antiferromagnetic temperatures
of the related Ising model, respectively. 
If $\Gamma_g$ is infinite dimensional in the large sense,
the critical inverse temperature of the spin glass model, $\beta_c$,
is given by
\begin{eqnarray}
\label{mapp}
\beta_c=\mathrm{min}\{\beta_c^{(SG)},\beta_c^{(F/AF)}\},
\end{eqnarray} 
where $\beta_c^{(SG)}$ and $\beta_c^{(F/AF)}$ are, respectively, 
the solutions of the equations, if any
%\numparts
\begin{eqnarray}
\label{mapp0}
\mediau{\tanh^2(\beta_c^{(SG)} J_b)}&=&w_{F}^{(I)}, \\
\label{mapp01}
\mediau{\tanh(\beta_c^{(F/AF)} J_b)}&=&w_{F/AF}^{(I)},
\end{eqnarray} 
where $F$ or $AF$, in the l.h.s and r.h.s. of Eq. (\ref{mapp01}), 
are in correspondence.  

\subsection{Generalization}
In the case of an arbitrary 
(not uniform) measure $d\mu_b$, useful for example for anisotropic
models, in which we have to consider even a given number of bond dependencies,
the related Ising model is defined by a set of,
typically few, independent couplings $\{J_b^{(I)}\}$, 
trough Eqs. (\ref{JI}-\ref{JIb})
and its critical behavior will be fully characterized 
by the solutions of an equation of the type $G_I(\{z_b\})=0$.
If $\Gamma_g$ is infinite dimensional in the large sense, 
Eqs. (\ref{mapp}-\ref{mapp01}) are generalized as follows.
The critical inverse temperature of the spin glass model $\beta_c$ is given by
\begin{eqnarray}
\label{mappg}
\beta_c=\mathrm{min}\{\beta_c^{(SG)},\beta_c^{(F/AF)}\},
\end{eqnarray} 
where $\beta_c^{(SG)}$ and $\beta_c^{(F/AF)}$ are, respectively, 
solutions of the two set of equations 
\begin{eqnarray}
\label{mapp0g}
G_I\left(\{\media{\tanh^2(\beta_c^{(SG)} J_b)}\}\right)&=& 0, \\
\label{mapp01g}
G_I\left(\{\media{\tanh(\beta_c^{(F/AF)} J_b)}\}\right)&=& 0,
\end{eqnarray} 

\subsection{Phase diagram}
Equations (\ref{mapp}-\ref{mapp0}) and their generalizations 
(\ref{mappg}-\ref{mapp01g}) give the
exact critical paramagnetic-spin glass ($P-SG$), $\beta_c^{(SG)}$, and
paramagnetic-$F/AF$ ($P-F/AF$), 
$\beta_c^{(F/AF)}$, surfaces, the
stability of which depend on which of the two ones is the minimum. 
In the case of a uniform measure, the suffix $F$ and $AF$ stay
for ferromagnetic and antiferromagnetic, respectively.
In the general case, such a distinction is not possible
and the symbol $F/AF$ stresses only that the transition is not $P-SG$.

\section{Spin glass over generalized tree-like graphs (static graphs)}
In the part I of this work \cite{I}, as application
of the mapping (\ref{mapp0}-\ref{mapp01}),
starting from the critical value of the universal parameter 
of the Ising model over a Bethe lattice of coordination number $q$, 
\begin{eqnarray}
\label{MF35}
w_F^{(I)}=\tanh\left(\beta_c^{(I)}J^{(I)}\right)=\frac{1}{q-1},
\end{eqnarray} 
we have recovered the critical surface
of the corresponding Ising spin glass model: a spin glass ($SG$) 
and a disordered ferromagnetic ($F$)
transitions take place at $\beta_c^{(SG)}$ and $\beta_c^{(F)}$, respectively 
solutions of the two following equations
%\numparts
\begin{eqnarray}
\label{MF37}
\mediau{\tanh^2\left(\beta_c^{(SG)}J_b\right)}=\frac{1}{q-1}, \\
\label{MF38}
\mediau{\tanh\left(\beta_c^{(F)}J_b\right)}=\frac{1}{q-1},
\end{eqnarray}
%\endnumparts
which, by using Eq. (\ref{mapp}), give
the upper phase boundary.

\begin{figure}[t]
\centering
\includegraphics[width=0.5\columnwidth,clip]{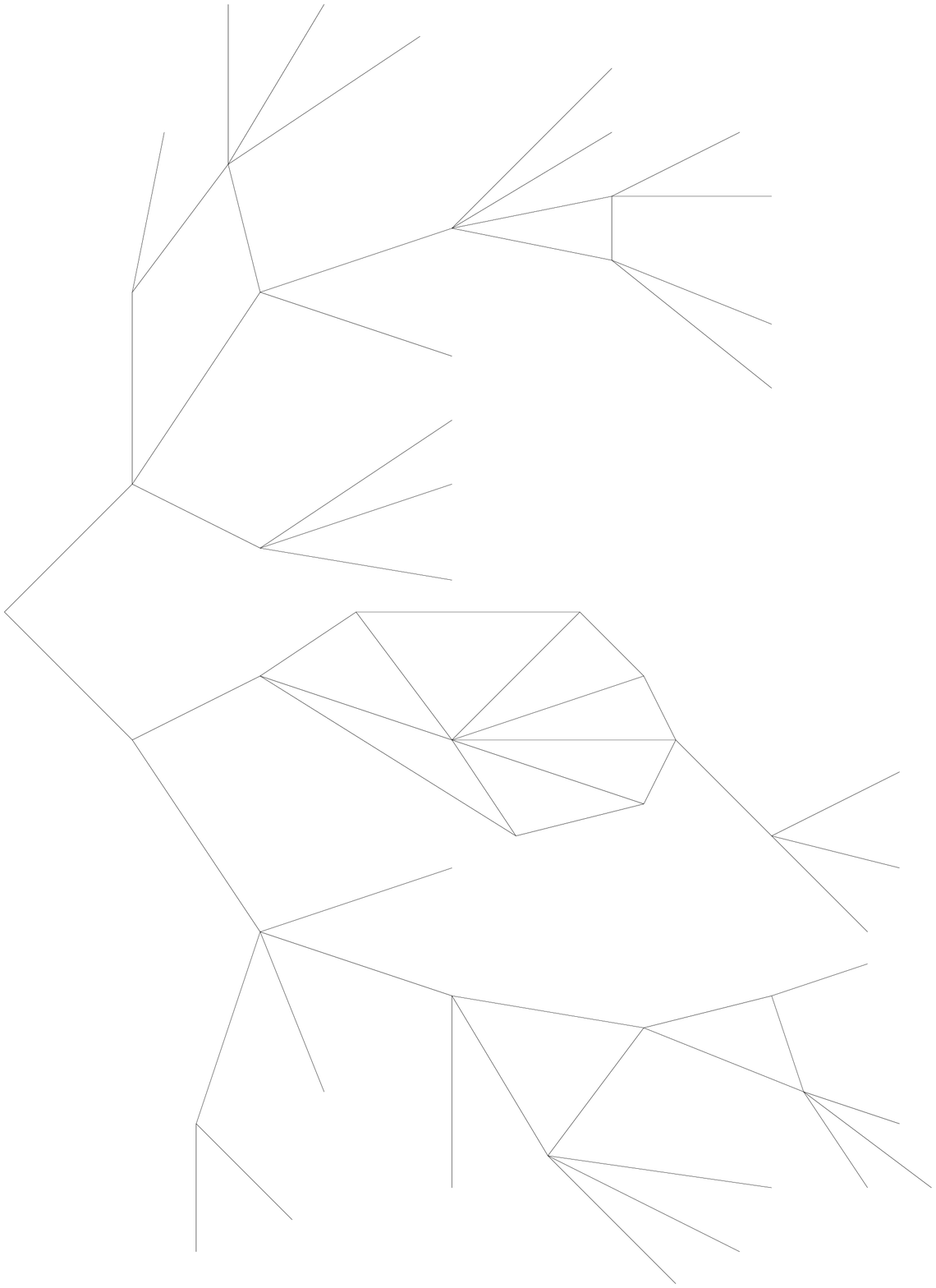}
\caption{ 
An example of a generalized tree-like graph with several loops.}
\label{tree}
\end{figure}

We now generalize Eqs. (\ref{MF37}) and (\ref{MF38}) to
a spin glass defined over a given generalized tree-like graph $g$. 
Here $g$ is said to be a generalized tree-like graph if for any vertex 
there is at most a finite number of loops (see Fig. \ref{tree}).
This application provides a new non trivial result. 
We will exploit the concepts and the conclusions of R. Lyons \cite{Lyons}.

Let us choose a vertex as the root $0$. If $v$ is a vertex, 
$|v|$ will indicate the number of links of the shortest path
connecting $0$ to $v$. A \textit{cutset} $\Pi$ will indicate
a finite set of vertices having the property that every 
infinite path starting from $0$ intersects $\Pi$.
In other words a cutset generalizes the concept of shell 
used in the Bethe lattice. Finally, to generalize the concept of
branching,
the \textit{branching number of} $\Gamma_g$, denoted by 
$\mathrm{br}~\Gamma_g$,
is introduced:
\begin{eqnarray}
\label{MF39}
\mathrm{br}~\Gamma_g\equiv
\inf\left\{\lambda>0:\inf_{\Pi}\sum_{v\in \Pi}\lambda^{-|v|}=0\right\}.
\end{eqnarray}
It is easy to check that in the case of a regular
lattice of coordination number $q$, one recovers $\mathrm{br}~\Gamma_g=q-1$.
Note however that, in general, $\mathrm{br}~\Gamma_g$ does not coincide
with the average of the number of branches per vertex. 

In \cite{Lyons} is proved that in the uniform Ising model with
positive coupling $J^{(I)}$ defined over a generalized tree like-graph,
the ferromagnetic phase transition occurs at 
\begin{eqnarray}
\label{MF40}
w_F^{(I)}=\tanh\left(\beta_c^{(I)}J^{(I)}\right)=
\frac{1}{\mathrm{br}~\Gamma_g},
\end{eqnarray}
which generalizes Eq. (\ref{MF35}). % of the regular Bethe lattice.
Correspondingly, by observing that a generalized tree-like graph is
infinite dimensional in the large sense (see Sec. 3 of \cite{I}), and 
by using the general rule (\ref{mapp0}-\ref{mapp01}) holding
for an Ising spin glass with a uniform measure, $d\mu_b\equiv d\mu$,
we immediately find 
that, a spin glass and a disordered ferromagnetic 
transitions, take place at $\beta_c^{(SG)}$ and $\beta_c^{(F)}$, respectively 
solutions of the two following equations
%\numparts
\begin{eqnarray}
\label{MF41}
\mediau{\tanh^2\left(\beta_c^{(SG)}J_b\right)}&=&
\frac{1}{\mathrm{br}~\Gamma_g}, \\
\label{MF42}
\mediau{\tanh\left(\beta_c^{(F)}J_b\right)}&=&\frac{1}{\mathrm{br}~\Gamma_g}.
\end{eqnarray}
%\endnumparts

Equations (\ref{MF41}) and (\ref{MF42}) 
generalize (\ref{MF37}) and (\ref{MF38}) to tree-like structures.

We can generalize even more the above result as follows.
In \cite{Lyons} is proved that for an arbitrary,
\textit{i.e.} non uniform, Ising model with positive couplings $\{J_b^{(I)}\}$ 
defined over a generalized tree like-graph, 
the ferromagnetic phase transition occurs  
at a critical temperature $T_c^{(I)}$ given by
\begin{eqnarray}
\label{MF43}
T_c^{(I)}=
\inf\left\{T:\inf_{\Pi}\sum_{v\in \Pi}\prod_{~\tau\leq v}
\tanh(J_{b(\tau)}/T)=0\right\},
\end{eqnarray}
where, for any vertex $v\in\Pi$, the product in the r.h.s. 
is extended to all vertices $\tau$ belonging to the
shortest path from $0$ to $v$ ($\tau\leq v$),
and $b(\tau)$ denotes the link connecting $\tau$ and its previous
point along the shortest path for $v$.
Formally, in terms of the universal quantities 
$\{z_b^{(I)}\}=\{\tanh(\beta J_b)\}$, Eq. (\ref{MF43}) says that 
the critical point $\{w_b^{(I)}\}$ of this non uniform 
Ising model is located at
\begin{eqnarray}
\label{MF44}
\{w_b^{(I)}\}=%\left(
\sup\left\{ \{z_b\}:
\inf_{\Pi}\sum_{v\in\Pi}\prod_{~\tau\leq v} z_{b(\tau)}=0 \right\}.
%\right)^{-1}.
\end{eqnarray} 

We can now, correspondingly, apply the rules (\ref{mapp0g}-\ref{mapp01g}) 
to a spin glass with a measure, $d\mathcal{P}=\prod_{b\in\Gamma_f} d\mu_b$,
and we find that a spin glass and a disordered ferromagnetic 
transitions take place at $\beta_c^{(SG)}$ and $\beta_c^{(F)}$, respectively 
solutions of the two following equations
%\numparts
\begin{eqnarray}
\label{MF45}
\beta_c^{(SG)}=
\sup\left\{ \beta:
\inf_{\Pi}\sum_{v\in\Pi}\prod_{~\tau\leq v} 
\int d\mu_{b(\tau)}\tanh^2(\beta J_{b(\tau)})=0 \right\},
\end{eqnarray} 
\begin{eqnarray}
\label{MF46}
\beta_c^{(F)}=
\sup\left\{ \beta:
\inf_{\Pi}\sum_{v\in\Pi}\prod_{~\tau\leq v} 
\int d\mu_{b(\tau)}\tanh(\beta J_{b(\tau)})=0 \right\}.
\end{eqnarray} 
%\endnumparts
Finally, the upper critical surface follows by applying Eq. (\ref{mapp}),
whereas the multicritical point $P-F-SG$ is the simultaneous 
solution of Eqs. (\ref{MF45}) and (\ref{MF46}) (system of equations).

Note that, in the Ref. \cite{Lyons}, Eqs. (\ref{MF42}) and (\ref{MF46}), 
for a disordered ferromagnetic Ising model, 
were already found but limited to
measures $d\mu_b$ with positive support of the couplings, $\{J_b>0\}$,
whereas Eqs. (\ref{MF41}) and (\ref{MF45}) for the spin glass
boundary, were still unpublished.

%Finally, we note that the relations (\ref{39}) and (\ref{40})
%turns out in close analogy with our mapping.

\section{Random graphs}
In the case of random graphs, an ensemble of graphs $G$, 
together with a probability distribution $P(g)$
for the elements $g$ of $G$, are given. The free energy of a spin glass 
model defined over the ensemble $G$ is then evaluated as
\begin{eqnarray}
\label{logZG}
-\beta F\equiv \sum_{g\in G}P(g)\int d\mathcal{P}\left(\{J_b\}\right)
\log\left(Z_g\left(\{J_b\}\right)\right),
\end{eqnarray} 
where $Z_g\left(\{J_b\}\right)$ 
is the partition function of the quenched system in the graph $g$
with couplings $\{J_b\}$.

In the following, we will apply explicitly the mapping to 
the so called ``equilibrium random graphs''
which are graphs maximally random under
the constraint that the degree distribution of a vertex is a given one, $p(k)$;
$k$ being the number of connections of the vertex.
For this family of graphs (but not only) the corresponding sets of links
$\Gamma_g$ are infinite dimensional in the large sense.
In fact, even if the number of infinitely length loops per vertex is infinite 
(see, \textit{g.e.}, \cite{Ginestra} and \cite{Marinari}), 
due to the fact that the bonds
are randomly distributed with the only constrain that they have to generate
the given degree distribution $p(k)$,
the probability that two of these loops overlap
each other for an infinite number of bonds is zero.
More in general, the elements of any ensemble $G$
are infinite dimensional in the large sense under the very weak condition
for the distribution $P(g)$ that the bonds are, each other, 
independent random variables.
Rather, the fact that the number of finite loops is infinite 
in certain scale free networks, has been
recently raised as a possible doubt about the validity of the
random tree approximation for the solution of the Ising model
on the equilibrium random graphs \cite{Ginestra} that we are about
to recall in the following subsection.
%Furthermore, only for such random graphs 
%the Ising model has been analytically solved.

\subsection{Ising model over equilibrium random graphs} 
Before to consider the Ising spin glass over a random graph,
we need to recall the general solution for the critical temperature
of the Ising model over equilibrium random graphs \cite{Goltsev,Zecchina}.
Let us indicate with $\mediaT{f(k)}_p\equiv \sum_{k=0}^\infty p(k)f(k)$
the average of the function degree $f(k)$ with respect to the degree 
distribution $p(k)$.  
The critical inverse temperature $\beta_c$ 
of the uniform Ising model with a coupling $J$ is given by
the following relation
\begin{eqnarray}
\label{TIG}
\tanh(\beta_cJ)=\frac{\mediaT{k}_p}{\mediaT{k^2}_p-\mediaT{k}_p}.
\end{eqnarray} 
It is interesting to see that this result can be approximately derived
even by using directly our mapping. 
%To this aim, let us consider the classical random graph of 
%Erd\ddot{o}s and R\acute{e}nyi \cite{Classical}. 
%In this ensemble, given $N$ vertices,
%$L$ bonds are uniformly distributed between the couple of vertices,
%giving rise to an average degree $\mediaT{k}=2L/N$.
%In this random graph, a given couple of vertices among a total of $N(N-1)/2$ 
%couples, will be connected by one of the $L$ bonds with
%a probability $p=2L/(N(N-1)$. Therefore, if in the ensemble 
%the correlations between the bonds are negligible, an Ising model
%with couling $J^{(I)}$ defined over the calssical random graph,
%will have a distribution of bonds given by
Let us consider a random graph with $N$ vertices and $L$ bonds so that
the average degree is $\mediaT{k}_p=2L/N$.
A given couple of vertices among the total of $N(N-1)/2$ 
couples, will be connected by one of the $L$ bonds with
a probability $p=2L/(N(N-1))=\mediaT{k}_p/(N-1)$. Therefore, if we neglect the
correlations among the bonds, an Ising model
with coupling $J$ defined over the graph
will have an effective distribution of couplings described by
\begin{eqnarray}
\label{dPG}
d\mathcal{P}\left(\{J_b\}\right)\equiv \prod_{b\in\Gamma_f} 
d\mu_b\left( J_b \right),
\end{eqnarray}
where
\begin{eqnarray}
\label{dmG}
d\mu\left( J_b \right) \equiv \delta(J_b-J)p+\delta(J_b)(1-p).
\end{eqnarray}
According to Eqs. (\ref{HI}-\ref{JIb}),
the related Ising model of this disorder model, 
whose disorder is given by Eqs. (\ref{dPG}) and (\ref{dmG}),    
is the Ising model defined over
the fully connected graph $\Gamma_f$ 
for which a ferromagnetic or antiferromagnetic
phase transition occurs at $\beta_c^{(I)}J^{(I)}N=1$; $J^{(I)}$ being
a generic coupling, positive or negative, respectively.
Therefore, if we now use the measure (\ref{dmG}) in Eq. (\ref{mapp01}), 
and $w_{F/AF}^{(I)}=\tanh(\beta_c^{(I)}J^{(I)})\to 1/N$ for $N$ large, 
we find that the critical inverse temperature $\beta_c$ 
of the Ising system over
the random graph, is given by $\tanh(\beta_c J)=1/\mediaT{k}_p$, 
which equals Eq. (\ref{TIG}) only in the case of the classical random graph,
for which $\mediaT{k^2}_p-\mediaT{k}_p=\mediaT{k}_p^2$ \cite{Classical}. 
The error is due to having neglected the correlations among the bonds
or, in other words, in assuming for $d\mathcal{P}(\{J_b\})$ a product measure. 
On the other hand, in a random graph,
the distribution of the bonds do not follow the
degree distribution of the vertices.
In fact, as is known, 
the probability that a bond points to a node of degree $k$
is not $p(k)$, but $p(k)k/\mediaT{k}_p$.
Correspondingly, one expects that, as an effective calculation,
in Eq. (\ref{dmG}) the degree average of a vertex $\mediaT{k}_p$, 
should be substituted with the degree average of the end of a bond:
$\mediaT{k^2}_p/\mediaT{k}_p$. By implementing this substitution one arrives
at $\tanh(\beta_c^{(I)}J^{(I)})=\mediaT{k}_p/\mediaT{k^2}_p$,
which constitutes a good approximation to the exact one 
(\ref{TIG}) 
%~\footnote{Such an intuitive argument has been just used in
%the Refs. \cite{Goltsev} and \cite{Zecchina}, 
%though within different procedures.}.
\cite{Comment}.

\subsection{Spin glass over random graphs}
As pointed out above, in a random graph it is not possible to
use an effective product measure $d\mathcal{P}(\{J_b\})$ which takes into
account rigorously also of the degree distribution.
However, for an exact treatment we can proceed as follows.
Let be given some disorder $d\mu(J_b)$ (we consider here only the
case of a same disorder for any bond).
We observe that, in Eq. (\ref{logZG}), the averages with respect to
$d\mathcal{P}(\{J_b\})$ and to $P(g)$ are interchangeable, so that,
if for any fixed graph $g$ we apply our mapping, (see Eqs. (67) 
of Ref. \cite{I}), for the free energy $F$ we find
\begin{eqnarray}\fl
\label{logZGS}
-\beta F&=&N\log(2)+
\sum_{g\in G}P(g)\sum_{b\in\Gamma_g}
\int d\mu_b(J_b)\log(\cos(\beta J_b)) %\nonumber \\ && 
+ N\sum_{g\in G}P(g)\varphi(g),
\end{eqnarray} 
where $\varphi(g)$ is proportional to the non trivial part of the 
high temperature expansion of the free energy density over the graph $g$ and,
at least in the limit $\beta\to\beta_c^{-}(g)$, it can be evaluated 
by $\varphi_{eff}(g)$ through the mapping equations (24-25) of Ref. \cite{I},
$\beta_c(g)$ being the critical inverse temperature of the spin glass model
defined over the (static) graph $g$ and has to be determined
according to Eqs. (\ref{mapp}-\ref{mapp01}).
In other words, for any fixed graph $g$, the mapping says that 
the critical behavior of the system is encoded
in the effective density $\varphi_{eff}(g)$ which
%according to Eqs. (24) and (25) of Ref. \cite{I}, 
is nothing else that the 
non trivial part of the high temperature expansion of the uniform Ising model
defined over the graph $g$ and in which, for any $b\in\Gamma_g$, 
the bond parameter $\tanh(\beta J)$ is simply substituted by 
$\int d\mu(J_b)\tanh(\beta J_b)$ or $\int d\mu(J_b)\tanh^2(\beta J_b)$,
depending on which transition we are interested in: disordered ferromagnetic or
spin glass, respectively.
%In other words, in the limit $\beta\to\beta_c^{-}(g)$, the mapping
%allows us to read effectively a given Ising spin glass model over a
%graph $g$ as an Ising model over $g$ in which we implement the substitutions
%$\tanh(\beta J)\to\int d\mu(J_b)\tanh(\beta J_b)$ or 
%$\tanh(\beta J)\to\int d\mu(J_b)\tanh^2(\beta J_b)$.
Since this is true for almost any graph $g\in G$ (with respect to the
measure $P(g)$),
we have that, in the limit $\beta\to\beta_c^{-}(g)$, 
the ensemble $G$ of the Ising spin glass models can be effectively read
as an ensemble $G$ of uniform Ising models in which the parameters
of the high temperature expansion are given by  
$\int d\mu(J_b)\tanh(\beta J_b)$ or $\int d\mu(J_b)\tanh^2(\beta J_b)$,
depending on which transition we are interested in.
Due to this Ising-representation, by using Eq. (\ref{TIG}) 
(which holds for an ensemble of uniform Ising models and 
where now $\beta_c$ and $J$ must be meant as quantities of the
related Ising model, \textit{i.e.},
$\beta_c^{(I)}$ and $J^{(I)}$, respectively) in 
Eqs. (\ref{mapp0}) and (\ref{mapp01}) applied to 
an ensemble of random graphs with weights $P(g)$ and
bond parameters $\int d\mu(J_b)\tanh(\beta J_b)$ or 
$\int d\mu(J_b)\tanh^2(\beta J_b)$, we find immediatly the following 
disordered ferromagnetic or spin glass critical temperatures, respectively
\begin{eqnarray}
\label{TIGS}
\int d\mu(J_b)\tanh(\beta_cJ_b)&=&
\frac{\mediaT{k}_p}{\mediaT{k^2}_p-\mediaT{k}_p}, \\
\label{TIGSb}
\int d\mu(J_b)\tanh^2(\beta_cJ_b)&=&
\frac{\mediaT{k}_p}{\mediaT{k^2}_p-\mediaT{k}_p}.
\end{eqnarray} 
Finally, by applying Eq. (\ref{mapp}), the upper critical 
surface of the phase diagram follows, whereas
the multicritical point $P-F-SG$ is the simultaneous solution of 
Eqs. (\ref{TIGS}) and (\ref{TIGSb}) (system of equations).
 
Equations (\ref{TIGS}-\ref{TIGSb}), 
together with their consequences, have been just derived in
the Ref. \cite{Kim} in the framework of the replica approach,
where an analysis of the order parameters has also been
done. We stress however that, unlike \cite{Kim}, where
the random graphs considered were confined to the particular class 
generated by the so called ``static model'' \cite{Goh}, our derivation
shows that Eqs. (\ref{TIGS}) and (\ref{TIGSb}) 
hold for the most general equilibrium random graph. 
More in general, by using the same above scheme, 
given an ensemble of random graphs $G$ distributed 
with some $P(g)$, if 
the set $\Gamma_g$ is infinite dimensional in the large sense
for almost any $g\in G$ with respect to $P(g)$, as
happens when the bonds are independent random variables,
the critical disordered ferromagnetic 
and spin glass critical temperatures are 
the solutions of the two following equations, respectively
\begin{eqnarray}
\label{TIGSG}
\int d\mu(J_b)\tanh(\beta_cJ_b)&=&w^{(I)}(G;P) \\
\label{TIGSGb}
\int d\mu(J_b)\tanh^2(\beta_cJ_b)&=&w^{(I)}(G;P),
\end{eqnarray} 
where $w^{(I)}(G;P)$ is the (universal) Ising critical parameter 
in the ensemble $G$ with distribution $P$ of the uniform Ising model
having coupling $J_b^{(I)}$;
$w^{(I)}(G;P)=\tanh(\beta_c^{(I)}(G;P) J^{(I)})$,
$\beta_c^{(I)}(G;P)$ being the inverse critical temperature. 
In the case of equilibrium random graphs, Eqs. (\ref{TIGSG})
and (\ref{TIGSGb}) reduce to Eqs. (\ref{TIGS}) and (\ref{TIGSb}),
respectively. 

As anticipated in the introduction, we stress that, in general, 
a possible direct numerical experiment
to measure the Ising critical parameter $w^{(I)}(G;P)$, turns out to be
hugely easier rather than a direct measure for the critical
temperature $\beta_c$ of the spin glass model.

\section{Conclusions}
A general mapping presented in the part I of this work \cite{I},
allows to derive the upper critical surface of an Ising spin glass
model starting from the critical temperature of a related
Ising model. When the dimensionality of the set of links $\Gamma$ is
infinite, the mapping is exact. 
In the present work, we have applied such a mapping to the case
of static generalized tree-like structures and networks.

Since in a generalized tree-like structure the number of loops
per vertex is finite, these structures turns out to be infinite
dimensional. Therefore, by using the known solution for the Ising
model over a generalized tree-like structure, we have derived the
upper critical surface. We stress that, the result found for the
ferromagnetic disordered transition, holds in the most general case, 
regardless of the measure $d\mu_b$ of the disorder,  
whereas the formulae for the spin glass transition were still unpublished.

Concerning networks, which are ensembles of random graphs 
characterized by some distribution $P(g)$,
the dimensionality is infinite as far as $P(g)$ does not generate 
infinite long paths with an infinite overlapping of bonds,
as happens in the so called equilibrium random graphs and,
more in general, when $P(g)$ is such that 
the bonds turns out to be each other independent random variables.
For the case of equilibrium random graphs, whose $P(g)$ can be completely
described in terms of the only degree distribution of the vertices 
$p(k)$, the known solution
for the Ising model allowed us to find exactly   
the upper critical surface of the Ising spin glass model over the 
network. In this case, as in the Ising model, the critical surface
is expressed in terms of the first and second moment of $p(k)$.
We stress that this result is very general, holding
for any kind of equilibrium random graph and not only for the so called
``static graph'' \cite{Kim}. 
Yet, we have provided a more general formula which, in terms
of the critical temperature of the Ising model, applies exactly
to any network as far as almost (with respect to the measure $P(g)$) 
any graph $g$ is infinite dimensional 
(as happens when the bonds are each other independent random variables).
Unlike the case of the equilibrium random graphs, however, 
now the critical temperature of the Ising model, and, therefore,
of the Ising spin glass model as well, will be expressed in terms of 
more complicated averages than the simple
moments of the degree distribution $p(k)$.
We point out however, that, 
even if in these cases the analytical knowledge of
the critical temperature of the Ising model is absent, 
a direct numerical measure for
it, turns out to be hugely easier than a direct measure for the critical
temperature of the spin glass model.

By looking back at: Eqs. (\ref{MF37}) and (\ref{MF38}) 
for the Bethe lattice case; Eqs. (\ref{MF41}) and (\ref{MF42}) 
or (\ref{MF45}) and (\ref{MF46}) for the generalized tree-like graphs;
Eqs. (\ref{TIGS}) and (\ref{TIGSb}) for the equilibrium random graphs;
and Eqs. (\ref{TIGSG}) and (\ref{TIGSGb}) for more general random graphs,
we recognize a sort of common structure for the critical surface and
the critical behavior. On the other hand, this is not surprising
considering that the related Ising models
of these models are essentially reduced to an Ising model over some
tree-like graph, if necessary, distributed according to some distribution
whose fluctuations has to be taken into account.
In this sense, we can say that the mapping establishes a sort
of universality among different Ising spin glass models.

We conclude by observing that, as already mentioned in the part I, 
our mapping does not seem to be peculiar of the Ising spin glass models. 
For more general models, a mapping at high temperature 
``random system'' $\rightarrow$ ``non random system'' seems possible;
so that a known solution of a given non random model can be
used to find the upper critical surface of the corresponding random model
over infinite dimensional graphs or infinite dimensional networks.

\section*{Acknowledgments}
%We thank ...%for S. Franz for useful discussions 
%and a critical reading of the manuscript.
This work was supported by DYSONET under NEST/Pathfinder initiative FP6, 
and by the FCT (Portugal) grant SFRH/BPD/24214/2005. 
%The research was also partially supported by Italian MIUR under 
%PRIN 2004028108$\_$001. 
I am grateful to A. V. Goltsev for useful discussions 
and to F. Mukhamedov for bringing Ref. \cite{Lyons} to my attention.
%Cofinanziamento MIUR 
%protocollo 2002027798$\_$001.


\begin{thebibliography}{99}
\bibitem{Comment} Such an intuitive argument has been just used in
the Refs. \cite{Goltsev} and \cite{Zecchina}, though within different
procedures.
\end{thebibliography}
\end{document}